\documentclass[conference]{IEEEtran}
\IEEEoverridecommandlockouts

\usepackage{cite}
\usepackage{amsmath,amssymb,amsfonts}
\usepackage{algorithmic}
\usepackage{graphicx}
\usepackage{textcomp}
\usepackage{xcolor}
\usepackage[caption=false]{subfig}

\def\BibTeX{{\rm B\kern-.05em{\sc i\kern-.025em b}\kern-.08em
    T\kern-.1667em\lower.7ex\hbox{E}\kern-.125emX}}
\begin{document}

\title{Device Invariance using Domain Adaptation on Acoustic Scene Classification}

\author{
\IEEEauthorblockN{Abhishek Dileep, Shubham Sharma, Padmanabhan Rajan}
\IEEEauthorblockA{School of Computing and Electrical Engineering\\
Indian Institute of Technology Mandi, Himachal Pradesh, India\\
Email: \{s23103, s23044\}@students.iitmandi.ac.in, padman@iitmandi.ac.in}
}

\maketitle

\begin{abstract}
This paper explores the effectiveness of domain adaptation techniques when using convolutional neural network (CNN)-based and transformer-based feature representations for acoustic scene classification. Two well-known domain adaptation techniques, namely domain adversarial neural network  (also called DANN) and conditional domain adversarial network (also called CDAN) are evaluated under various domain shifts. Our study indicates that DANN provides effective domain adaptation fairly consistently for both feature extractors. On the other hand, CDAN provides effective domain adaptation only for CNN-based feature extractors. The study gives insights into how domain adaptation methods may need to be tailored to the underlying feature representation. Experimental evaluation with multiple devices on the DCASE 2020 dataset supports the observations.
\end{abstract}


\section{Introduction}
Over the past decade, deep learning models have made significant progress in classification tasks. In acoustic scene classification (ASC), the task is to assign a single label (such as ``park'', ``metro station'', ``bus'' etc.) to an acoustic scene represented by an audio signal. The label represents the acoustic environment where the signal was recorded. The Detection and Classification of Acoustic Scenes and Events (DCASE) series of challenges\footnote{https://dcase.community/} have resulted in advancement in the field. Deep architectures such as convolutional neural networks (CNNs) as well as transformers have resulted in excellent classification performance \cite{rf_reg}, \cite{Passt}, \cite{CAA-Net}, \cite{DCASENET}.

However, many deep learning models can be poor at generalization when there is a domain shift. This problem has been well documented in computer vision. Here, if the training images come from one domain (e.g., front-facing face images), and the test images come from a different domain (e.g., oblique-facing images), there is a drop in classification performance. Similarly, for audio data, a domain shift results in a drop in classification performance. For example, if the training data has speech from adults, but if the test data has speech from children, there will be a drop in performance. Similarly, a domain shift occurs if the training data comes from one device, and the test data comes from another device. This has been shown to affect performance in ASC \cite{disentanglement}, \cite{ASC_with_ADA}, \cite{ASC_with_WGAN}, \cite{MTDA}.

Various methods for domain adaptation have been proposed in the literature to overcome domain shift \cite{domainAdaReview}. These methods include methods which are instance-based, feature-based, and which are adversarial in nature. The purpose of this paper is to investigate how domain adaptation methods perform with respect to different feature representations, particularly CNNs and transformers, in the context of ASC. In our study, domain shift is represented by change in the capturing device (training data is from device X and test data is from device Y.) 

The experimental evaluation in this paper compares the performance of two well-known domain adaptation methods, namely domain adversarial neural network also known as DANN and conditional domain adversarial network also called as CDAN. Our results suggest that adaptation methods may need to be crafted depending upon the underlying feature representation. We support our hypothesis with extensive experiments with multiple devices on the DCASE 2020 dataset for ASC.


\section{Related Works}
\subsection{Feature representations}
Recent methods for ASC include various processing pipelines which include CNN-based feature extractors such as \cite{CliqueNets}, \cite{CAA-Net}, and \cite{rf_reg}, and transformer-based feature extractors such as \cite{AST}, \cite{Passt} and \cite{M2D}. Works such as regularized receptive field CNNs \cite{rf_reg} have resulted in effective representations for audio for the ASC task. The DcaseNet \cite{DCASENET} is another CNN-based architecture that is jointly trained with other similar tasks like sound event detection and audio tagging using various DCASE datasets. Pre-trained networks such as OpenL3-Net\cite{L3net} have also been used for feature representations.

Self-attention-based models derived from vision transformer \cite{Vit} used in ASC include audio spectrogram transformer (AST) \cite{AST} and its derivatives such as 
patchout spectrogram transformer (PaSST) \cite{Passt} and masked modelling duo (M2D) \cite{M2D}. These models are pre-trained on AudioSet \cite{AudioSet}.

    \subsection{Acoustic Scene Classification - Domain Adaptation methods}

Domain adaptation methods used in the context of ASC have generally been derived from the existing literature. In particular, we focus on unsupervised domain adaptation, where there is labeled data available from one domain (the source domain) and unlabeled data available from another domain (the target domain.) These include methods such as domain adversarial neural networks (DANN) \cite{DANN} that use adversarial training along with a layer for reversing the gradient to learn representations that are agnostic to domain shift. Another similar method is the conditional domain adversarial network which conditions the adversarial training, resulting in domain invariant features. The adversarial discriminative domain adaptation method (ADDA) \cite{ADDA} uses separate encoders for target and source domains. Maximum classifier discrepancy (MCD) uses two classifiers and tries to detect the target samples far from the source support. Other methods for domain adaptation include the Gradually Vanishing Bridge (GVB) \cite{gvb}, which tries to reduce the domain-specific information using a bridge layer. Cycle Self-Training (CST), \cite{cst} uses cyclic training on pseudo-labels generated by the target classifier and enforces them on the source domain. 

Authors have explored domain adaptation in the context of ASC using devices as domains \cite{ASC_with_ADA}, \cite{ASC_with_WGAN}, \cite{MTDA} and \cite{BlindDA_with_BN}. Other works include the use of recording locations (cities) as domains \cite{Prototypical}, \cite{disentanglement}. In this paper, we focus on devices as the domain.

\section{Processing pipelines}
    In this section we explain the different feature extractors and domain adaptation methods we use in our experimental evaluation.
    
    \subsection{PaSST - Feature Extractor} 
        
        The PaSST (Patch-Spectrogram Transformer) model uses a transformer encoder-decoder architecture and takes inspiration from ViT's \cite{Vit} architecture to train. The model receives a patched spectrogram as input. These patches overlap with each other, effectively increasing the receptive field of the model while keeping the complexity relatively low. 

        The model learns better representations by dropping parts of the input sequence patches randomly. This forces the model to perform classification using incomplete patches similar to drop out layers. This is only done during pretraining time to make the PaSST more robust and mainly decrease the model computational complexity as described in \cite{Passt}. During fine-tuning as described in later sections, the whole spectrogram is passed through the model without patch dropping.  
        
        The structured patchouts are rows and columns of frequency/time bins and are picked at random and removed before passing to the model. A patch size of $16\times16$ (16 time frames and 16 mel frequency bins) are used. Unlike ViT which uses a 3-channel input to the patch embedding layer, the PaSST uses a single channel. 
        
        The spectrogram is magnitude normalized with a zero mean and 0.5 standard deviation. 
        The PaSST model is pre-trained for classification tasks using ImageNet and AudioSet \cite{AudioSet} datasets. During pretraining, the PaSST model uses 2M samples from Audioset for training and 20K samples for testing. 
        
\begin{figure} 
        \centering 
        \includegraphics[width=0.8\linewidth]{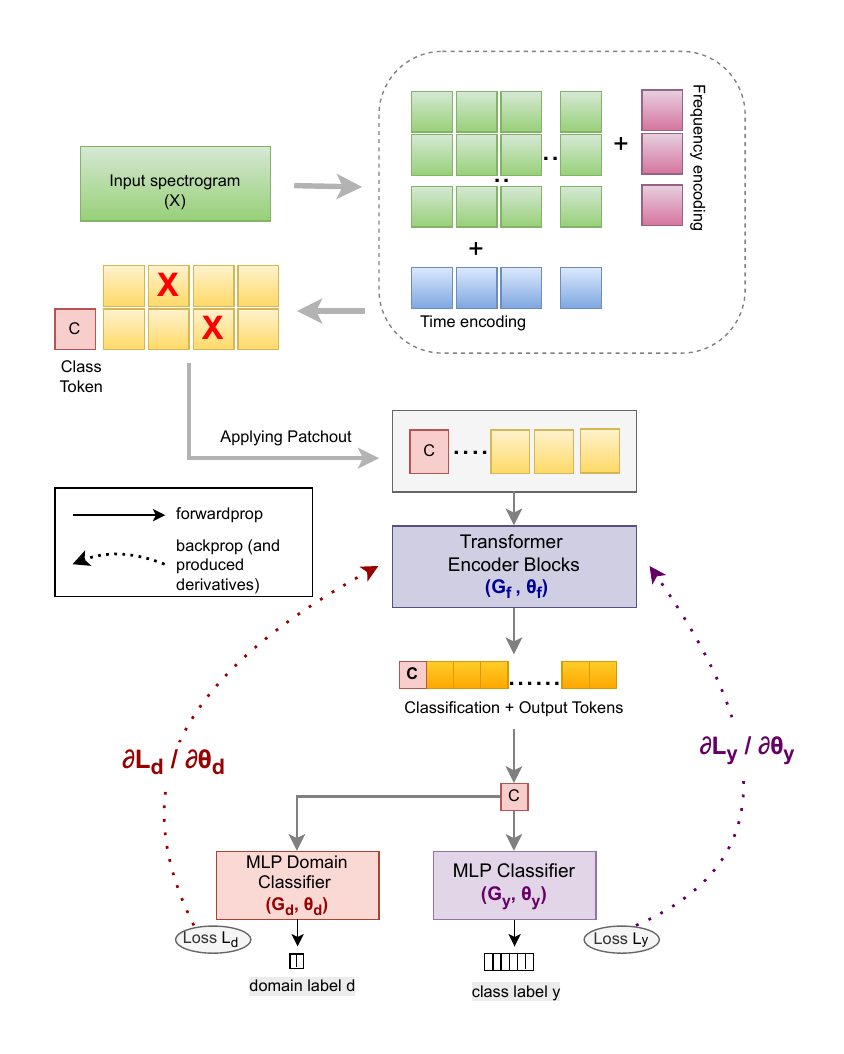} 
        \caption{PaSST model with classifier and domain discriminator. During fine-tuning, only purple blocks are updated. During domain adaptation, both purple and pink blocks are updated with their respective losses.} 
        \label{fig:passt_fig} 
        \end{figure}

        This pre-trained model is then fine-tuned as described in the Experiments section using the classification loss
        
        \begin{equation}
        \label{cross_entropy_loss}
        \mathcal{L}_{\mathrm{cls}} = \mathcal{L}_{ce}( G_{\theta_y} (G_{\theta_f} (\mathcal{X}_s)), \mathcal{Y}_s),    
        \end{equation} 
        where $L_{ce}$ denotes the cross-entropy loss. In Fig. \ref{fig:passt_fig} the cross entropy loss is denoted as $L_y$ which are calculated over class label $y$. 
        The classification token (CLS, denoted C in the Fig. \ref{fig:passt_fig}) is added to attend to global features.
        In our current setup, we use only the classification token for fine-tuning, while the sequential outputs are discarded. During fine tuning, the purple colored blocks in Fig. \ref{fig:passt_fig} get updated.
        
        
    \subsection{DcaseNet - Feature Extractor}
        The DcaseNet model proposes joint training of three audio-related tasks, namely ASC, audio tagging (TAG) and sound event detection (SED). 
        It considers the joint training of ASC, TAG and SED with a CNN backbone and uses gated recurrent units (GRU) layers for sequence modeling. 
        
\begin{figure} 
        \centering 
        \includegraphics[width=0.8\linewidth]{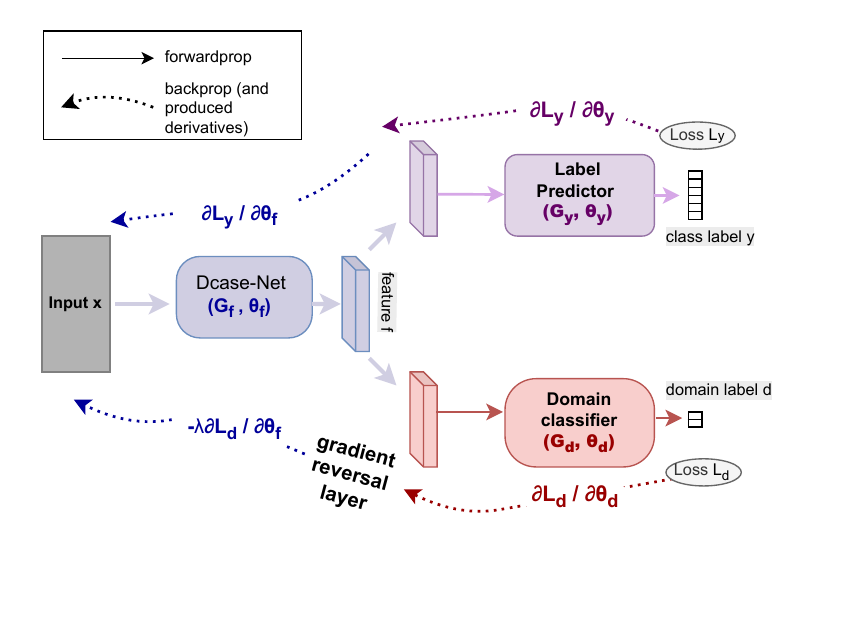} 
        \caption{ DcaseNet Model with Classifier and Domain Discriminator. During fine-tuning, only purple blocks are updated. For domain adaptation, both purple and pink blocks are updated with their respective losses.} 
        \label{fig:DcaseNet} 
        \end{figure}

        This model is pre-trained on multiple DCASE datasets for ASC, TAG, and SED and can be further fine-tuned for a specific task. In our case, we consider the output of the CNN model for fine-tuning, as shown in Fig. \ref{fig:DcaseNet}, where $L_y$ is the loss function $L_{cls}$ as shown in eqn \ref{cross_entropy_loss}. 
        
    \subsection{Custom CNN - Feature Extractor}
        This is a custom feature extractor using a two-layer convolutional neural network (CNN) that takes input spectrograms and extracts  local information from them. This architecture is not pre-trained and is trained from scratch. There are 32 filters with a kernel size of 7×7 in the first convolutional layer. After that, there is batch normalization, ReLU activation, max-pooling with a 5×5 window, and spatial dropout to improve generalization. Similarly, the second CNN layer has 64 filters and adds a dropout and max-pooling layer. The extracted feature maps are then flattened into a 1D final layer, which will be used as the learned representation. This is henceforth referred to as custom CNN in the rest of the paper.
    
    \subsection{DANN - Domain Adaptation}  
    We briefly describe the domain adversarial neural network (DANN.) The DANN uses adversarial training to learn features which are invariant to the source and target domains. This is done with a domain discriminator which tries to discriminate between source and target domains, and the feature extractor which produces features to confuse the discriminator. The loss function for DANN is given as \cite{DANN}:  
        
        \begin{align}
        \begin{split}
        \mathbb{E}(\theta_f, \theta_y, \theta_d) &= \sum_{(x_s,y_s) \sim D_s} L_y\big(G_y(G_f(x_i;\theta_f);\theta_y),y_i\big) \\
        &\quad - \lambda \sum_{(x_i , d_i) \sim (D_s \cup D_t ) } L_d\big(G_d(G_f(x_i;\theta_f);\theta_d),d_i\big)
        \end{split}
        \end{align}
        
        Here $\theta_f$, $\theta_y$, $\theta_d$ denote the parameters of the feature extractor $G_f$, classifier $G_y$ and domain discriminator $G_d$ respectively. $x_i$ is an input sample, $y_i$ is the target label and $d_i$ is the domain label, $L_y$ denotes the cross-entropy loss calculated over $(x_i^s, y_i^s) \sim D_s$ from the source domain, and $L_d$ is the binary cross-entropy loss calcualated over $(x_i , d_i) \sim (D_s \cup D_t)$. $d_i$ is the domain label. 

    The min-max optimization becomes        
        \begin{align}
        \label{opt_eqn}
        (\hat{\theta}_f, \hat{\theta}_y) &= \arg\min_{\theta_f, \theta_y} \mathbb{E}(\theta_f, \theta_y, \hat{\theta}_d), \\
        \hat{\theta}_d &= \arg\max_{\theta_d} \mathbb{E}(\hat{\theta}_f, \hat{\theta}_y, \theta_d).
        \end{align}
    where $\hat{\theta}_f, \hat{\theta}_y, \hat{\theta}_d$ delivers the saddle point of the function.  
    
    \subsection{CDAN - Domain Adaptation}
        Conditional Domain Adversarial Networks (CDAN) extend the idea of DANN by
        making the domain discriminator conditioned on both the feature representations and the classifier predictions \cite{CDAN}. 
        
        
        Similar to DANN, the CDAN framework consists of three components: the feature extractor,
        the classifier, and the domain discriminator. Instead of using 
        the feature representation alone, the discriminator operates on the
        outer product of the feature representation and the classifier predictions, 
        thus aligning the joint distribution of features and outputs across source 
        and target domains.
        
       The CDAN can be expressed as a min-max optimization problem expressed as 

        \begin{multline}
        \mathbb{E}(\theta_f,\theta_y,\theta_d) =
        \sum_{(x_s,y_s)\sim D_s} L_y\big(G_y(G_f(x_i;\theta_f);\theta_y),y_i\big) \\
        - \!\lambda \Big( \sum_{(x_i,d_i)\sim (D_s \cup D_t)} 
        L_d\big(G_d(G_f(T(h_i);\theta_f);\theta_d),d_i\big) \Big).
        \end{multline}

\begin{figure} 
        \centering 
        \includegraphics[width=0.8\linewidth]{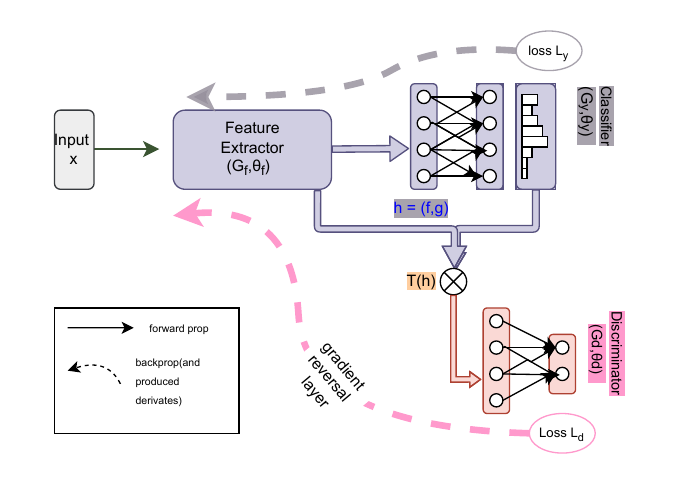} 
        \caption{CDAN for source device to target device. For baseline model, only purple blocks are used. For adaptation, purple and pink blocks are used with their respective losses.} 
        \label{fig:cdan_fig} 
        \end{figure}

        
        This equation is similar to DANN except the domain discriminator $G_d$ which is conditioned on $T(\cdot)$. $T(\cdot)$ is the outer product operation for the joint variable  $h_i = (f_i, g_i)$. The $f_i$ are features from the feature extractor and $g_i$ are class predictions from $G_y$ classifier. This is shown in Fig. \ref{fig:cdan_fig} where $L_d$ is binary cross entropy loss and $L_y$ is the classification loss. The min-max optimization is the same as DANN  equation \ref{opt_eqn}.

        There exists the issue of increased dimension to the input layer of the MLP 
        due to the multilinear output over which the discriminator has been conditioned, hence a 
        fixed-parameter random layer is used on the input which downsamples the input dimensions to 
        lower dimensions. This random layer does not affect the error bound of the model as proved in \cite{CDAN}.
        By conditioning the discriminator on the source and target predictions, CDAN provides a more discriminative alignment, 
        thus reducing the discrepancy between source domain and target domain more effectively than DANN. This leads to improved
        generalization performance on the target domain.
        
\section{Experiments}
In this part of the paper, we first explain the dataset in section A, the implementation details in section B. In section C we describe the baseline source only models, and subsequently section the domain adaptation methods applied in sections C and D. 

\subsection{Dataset}
We utilize the DCASE 2020 Task 1A dataset, which has 64 hours of data with 10 acoustic scene classes. Since our study specifically focuses on domain adaptation due to device mismatch, we do not follow the standard DCASE protocol in our experimental evaluation. We consider six devices present in the train and eval portion of the dataset, namely devices A, B, C, S1, S2, and S3. Here devices A, B, and C are real devices, and S1, S2, and S3 are simulated devices. The audio for the simulated devices is created by convolving the impulse response of real devices with the audio captured by device A.
Details of the dataset are displayed in Table \ref{tab:device_dataset}. The total amount of audio in the development set we used is 55 hours, of which 40 hours are from device A, and 3 hours each from the remaining devices. Audio was provided in single-channel 44.1 kHz 24-bit format. 

In our experiments, the source domain is device A, and the other devices are target domains. Specifically, the data of device A from the train partition is used to train the source classifier and the data of device A from the test partition is used as the seen test data (no domain mismatch, same device.) The data of devices C, B, S1, S3 and S2 from the train partition is used for domain adaptation using DANN and CDAN, along with data from device A using only source/target domain labels. The data of the test partition from devices B, C, S1, S2 and S3 is used as the unseen target data (with domain mismatch, devices different from device A.) Our unsupervised domain adaptation setup thus uses class labels for the source domain and uses domain labels for the source/target domain (no class labels are used in the target domain.) Hence our domain shift is captured by training on device A and testing on the other devices. Thus the notation $\textrm{Dev}A \rightarrow \textrm{Dev}X$ represents a domain shift from device A to device X.


\begin{table}[ht]
\centering
\caption{Device Dataset Cross-validation setup details from DCASE 2020 used in this paper.} 
\label{tab:device_dataset}
\begin{tabular}{|l|p{0.5cm}|p{0.8cm}|p{0.8cm}|p{0.8cm}|p{0.8cm}|p{0.8cm}}
\hline
\textbf{Name} & \textbf{Type} & \textbf{Total length} & \textbf{Total samples.} & \textbf{Train samples.} & \textbf{Test samples.}  \\
\hline
A & Real & 40h    & 14400    & 10215 & 330  \\
B, C & Real & 3h & 1080 & 750 & 330  \\
S1, S2, S3 & Sim. & 3h & 1080 & 750 & 330  \\
\hline
\end{tabular}
\end{table}

\subsection{Implementation details}
We use a 2-layer classifier that takes an input of 768 embeddings for the PaSST feature extractor. For the CNN models, we take 384 as the input for both DcaseNet and the custom CNN. Regarding DcaseNet, we use an adaptive average pooling layer to pool all the outputs from the CNN blocks. For the custom CNN, we have used a final flattening layer. 
For all the three feature extractors, a fully connected layer of 128 and an output layer for 10-class classification is used. 
The loss is backpropagated to the PaSST model through the classification token, and in the CNN models all the layer parameters are updated. We use the Adam optimizer with learning rates of $1e{-5}$ and $1e^{-3}$ for feature extractor and classifier/domain discriminator respectively . Our implementation uses PyTorch and codebase from the original authors, and was trained on an A5000 GPU.

\subsection{Baseline - Source only models}
To see the effect of domain shift, we evaluate classification performance in the matched setting (train and test from the same device) and in the mismatched setting (train from device A and test from another device.) Using the three feature extractors, namely PaSST, DcaseNet, and custom CNN, the baseline performance (i.e., in the matched setting) is provided in the first row of Tables \ref{tab:passt_dann_results}-\ref{tab:dcasenet_results}. It can be seen that PaSST with its pretraining and fine-tuning on large amounts of data, and its self-attention mechanism, provides the highest baseline accuracy (0.845). The DcaseNet, trained with data from the DCASE challenges, gives slightly poorer performance (0.711). The custom CNN model, trained from scratch using only data from device A, shows lowest baseline performance (0.612.) No data augmentation was used in our studies.

As can be seen from the column 3 (titled Base Acc.) of the remaining rows of tables \ref{tab:passt_dann_results}-\ref{tab:dcasenet_results}, there is a significant drop in accuracy for mismatched scenarios: PaSST test accuracy drops an average of 20.6\%, CNN drops 36.6\%, and DcaseNet drops 5\% accuracy. This demonstrates the effect of domain shift, with device A representing the source domain, and the other devices representing the target domain.

\subsection{Domain adaptation}

The result of applying domain adaptation is shown in columns 4 and 5 of tables \ref{tab:passt_dann_results}-\ref{tab:dcasenet_results}. 
When using DANN, the accuracy for PaSST features increases by 8.5\%, for custom CNN features by 7.04\%, and for the DcaseNet feature extractor by 10.9\% from the base accuracy. DANN adaptation works in most cases for all three feature extractors. In particular, the custom CNN feature extractor seems to struggle for effective domain adaptation using DANN for the simulated devices S1, S2 and S3, as has been documented in \cite{MTDA}.  

CDAN performs effective domain adaptation for both DcaseNet as well as for the Custom CNN, even for the simulated devices. 
When using CDAN, accuracy with custom CNN features increases by 11.8\% and with the DcaseNet feature extractor by 11.5\% from the base accuracy.
On the other hand, CDAN fails to converge effectively for the PaSST feature extractor (model fails to converge, last column of Table \ref{tab:passt_dann_results}).

\begin{figure}[htb] 
    \centering
    \subfloat[t-SNE plot for mismatched condition for PaSST features]{
    \includegraphics[width=0.48\textwidth]{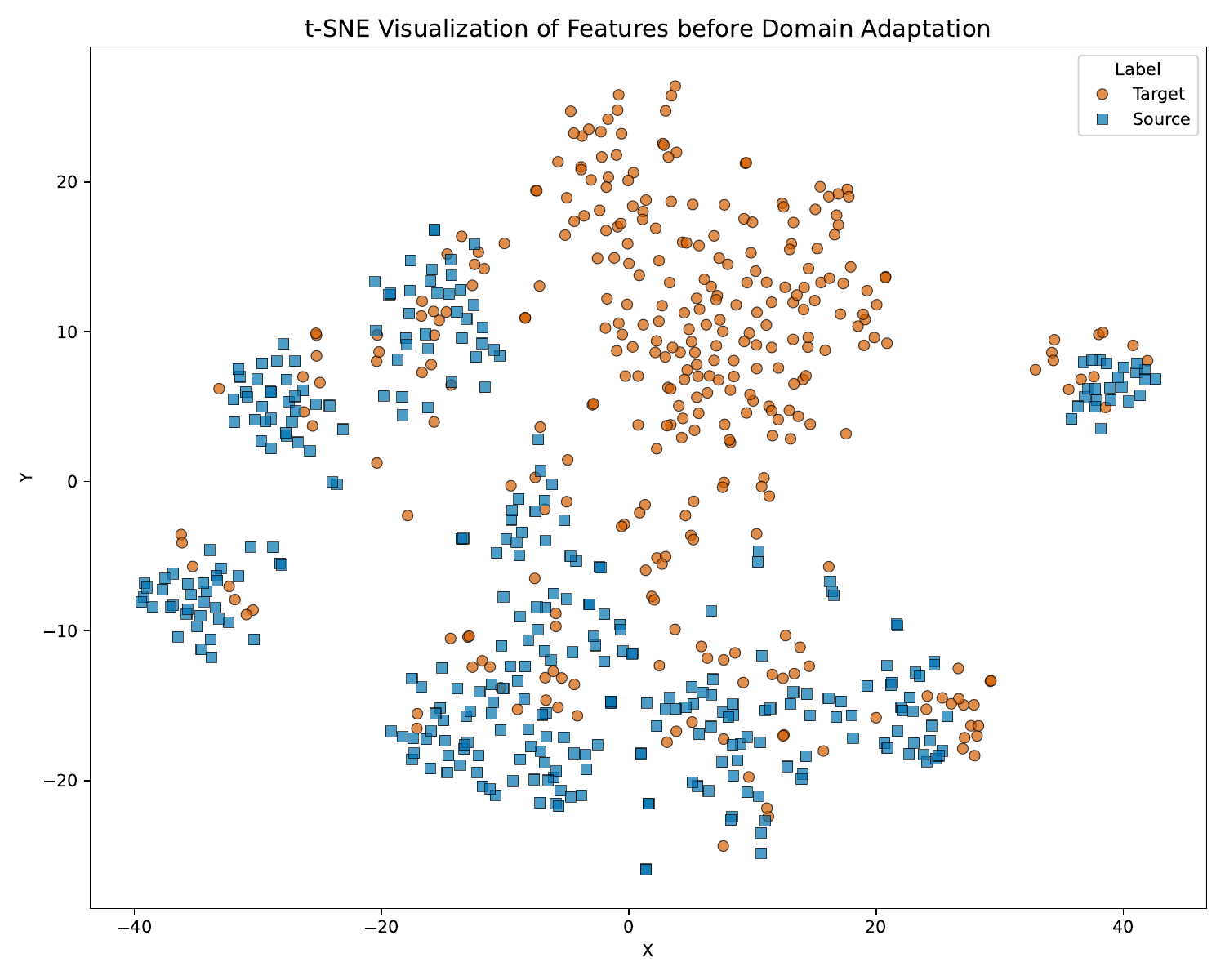}
    \label{fig:before}  
    }
    \hfill 
    \subfloat[t-SNE plot after applying DANN for PaSST features]{
    \includegraphics[width=0.48\textwidth]{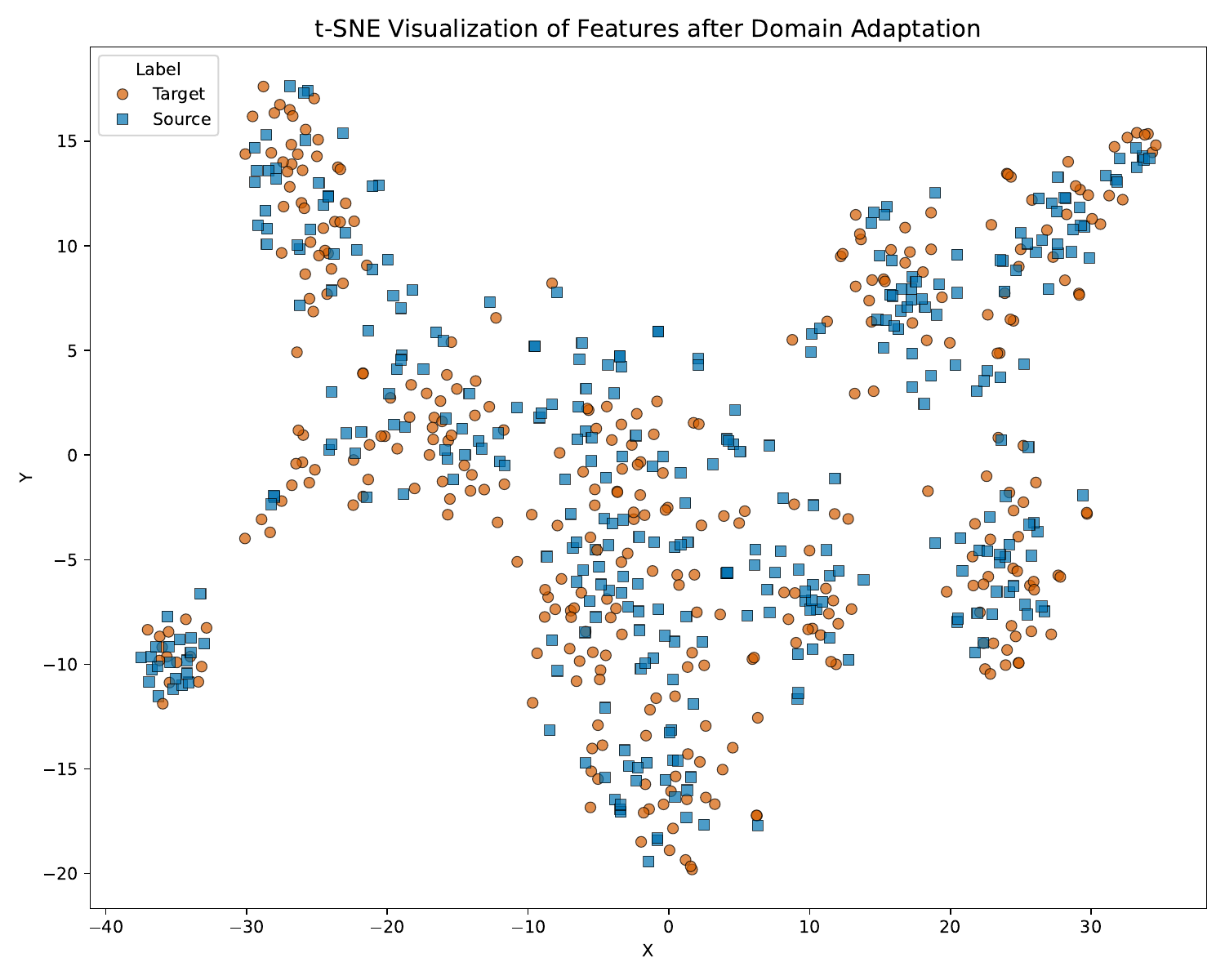}
    \label{fig:after}
    }
    \caption{Visualization and comparison of t-SNE plots for PaSST features: subplot \ref{fig:before}, \ref{fig:after}. Domain adaptation results in higher overlap between source and target domain data points.} 
    \label{fig:combined_tsne}
\end{figure}

We have generated t-SNE plots for PaSST features in Fig. \ref{fig:combined_tsne}. The t-SNE plot is calculated for 330 samples from source and target device chosen at random. The features are spread across all 10 classes. The plot shows that the spread has decreased for the PaSST features after applying DANN,.

\begin{table}[h]
\begin{tabular}{|p{1cm}|p{1.7cm}|p{1.5cm}|p{1.5cm}|p{1cm}|}
\hline
\textbf{Source Device} & \textbf{Target Device} & \textbf{Base Acc.} & \textbf{DANN } &\textbf{CDAN} \\
\hline
A   &  & 0.845    & -  & -   \\
A  & DevA$\rightarrow$DevB & 0.635    & 0.73 & -  \\
A  & DevA$\rightarrow$DevC & 0.650    & 0.74 &  - \\
A & DevA$\rightarrow$DevS1 & 0.645   & 0.65  & - \\
A & DevA$\rightarrow$DevS2 & 0.610   & 0.74  & - \\
A & DevA$\rightarrow$DevS3 & 0.655   & 0.74  & - \\
\hline
\end{tabular}
\caption{Accuracy of base PASST models and their performance after domain adaptation using DANN and CDAN.}
\label{tab:passt_dann_results}
\end{table}
\begin{table}[h]
\begin{tabular}{|p{1cm}|p{1.7cm}|p{1.5cm}|p{1.5cm}|p{1cm}|}
\hline
\textbf{Source Device} & \textbf{Target Device} & \textbf{Base Acc.} & \textbf{DANN } &\textbf{CDAN} \\
\hline
A   &  & 0.6121    & -  & -   \\
A  & DevA$\rightarrow$DevB & 0.243  & 0.386 & 0.33  \\
A  & DevA$\rightarrow$DevC & 0.246    & 0.55 & 0.48 \\
A & DevA$\rightarrow$DevS1 & 0.249   & 0.20  & 0.48 \\
A & DevA$\rightarrow$DevS2 & 0.237   & 0.18  & 0.26 \\
A & DevA$\rightarrow$DevS3 & 0.252   & 0.26  & 0.36 \\
\hline
\end{tabular}
\caption{Accuracy of custom CNN models and their performance after domain adaptation using DANN and CDAN.}
\label{tab:cnn_results}
\end{table}
\begin{table}[h]
\begin{tabular}{|p{1cm}|p{1.7cm}|p{1.5cm}|p{1.5cm}|p{1cm}|}
\hline
\textbf{Source Device } & \textbf{Target Device} & \textbf{Base Acc.} & \textbf{DANN } &\textbf{CDAN} \\
\hline
A   &  & 0.722   & -  & -   \\
A  & DevA$\rightarrow$DevB & 0.58    & 0.711 & 0.711  \\
A  & DevA$\rightarrow$DevC & 0.583    & 0.671 & 0.699 \\
A & DevA$\rightarrow$DevS1 & 0.545   & 0.72  & 0.718 \\
A & DevA$\rightarrow$DevS2 & 0.548   & 0.70  & 0.706 \\
A & DevA$\rightarrow$DevS3 & 0.578   & 0.684  & 0.684 \\
\hline
\end{tabular}
\caption{Accuracy of base DcaseNet models and their performance after domain adaptation using DANN and CDAN.}
\label{tab:dcasenet_results}
\end{table}


\section{Discussion}

   Our experiments indicate that effective methods for domain adaptation depend on the feature extraction architecture. The
    results also suggest that DANN improves on almost all feature extractors considered, including pre-trained PaSST, DcaseNet
    and custom CNN. 
    In contrast, CDAN, which performs well for CNN-based models, fails on the transformer-based PaSST. 
    We observed that mode collapse occurs in the classifier when trained with ViT feature extractor while using CDAN for domain adaptation.  
    CDAN relies heavily on pseudo labels provided by the classifier for conditioning the discriminator.
    On the other hand, classifiers trained on pseudo labels from CNN-based features don’t vary much due to the Gaussian nature of the effective receptive field and local inductive bias \cite{effecctiverf}.
In contrast ViT models contain a more global inductive bias \cite{transformerrf}, hence providing more dynamic features. Preliminary studies reveal differences in the statistical properties of these features. These will be the subject of future work.



\section{Conclusions}
In this work, we have explored the efficacy of popular domain adaptation methods with different feature extraction architectures. Our experiments on CNN and transformer based architectures revealed an important insight: The performance of a domain adaptation methods is dependent on the nature of feature extraction architecture. We demonstrated that DANN
works well in almost all conditions improving performance across all architectures, indicating that it is a robust and generalizable domain adaptation method. Meanwhile, CDAN seems to be more effective with CNN based feature extractors but fails to adapt with transformer based feature extractor like PaSST.
This suggests that we cannot blindly apply CNN based domain adaptation methods on transformer based feature extractors and need to careful evaluated for each context. 

Our analysis was confined to two specific adaptation methods and the DCASE 2020 dataset. The results may not generalize to all possible adversarial techniques or acoustic environments. In the future, we would like to explore more on the dynamics of domain adaptation with transformer based architectures and statistical properties of the features generated from these during adversarial training for domain adaptation.


\bibliographystyle{IEEEtran}
\bibliography{main_paper}

\end{document}